\documentclass[conference]{IEEEtran}
\IEEEoverridecommandlockouts 
\usepackage{siunitx}
\usepackage{cite}
\usepackage[hidelinks]{hyperref}
\usepackage{amsmath,amssymb,amsfonts}
\usepackage{graphicx}
\usepackage{textcomp}
\usepackage{xcolor}
\usepackage{booktabs}
\usepackage{multirow}
\usepackage{url}
\usepackage[section]{placeins}
\begin{document}

\title{\vspace{0.6in}\fontsize{19.2}{22.5}\selectfont
Reliability-Aware ETF Tail-Risk Monitoring}

\author{
\vspace{-5em}
Tenghan Zhong$^{1}$\thanks{$^{1}$Tenghan Zhong is with the Department of Mathematics, University of Southern California, Los Angeles, CA, USA (e-mail: tenghanz@usc.edu).}
\and
Keyuan Wu$^{2}$\thanks{$^{2}$Keyuan Wu is with the Department of Mathematics, University of Southern California, Los Angeles, CA, USA (e-mail: keyuanwu@usc.edu).}
}

\maketitle

\begin{abstract}
Daily ETF risk monitoring can become unreliable when market data quality degrades, market conditions shift, or predictive performance becomes unstable. This paper develops a reliability-aware risk monitoring service for next-day tail-risk surveillance. The proposed framework combines service-time quality checks, lower-tail prediction, uncertainty scoring, and risk-aware adjustment of the tail-risk estimate. We evaluate the system on a daily panel of multiple ETFs augmented with VIX and yield-curve information under a rolling walk-forward design. Empirically, the framework improves tail-risk monitoring, especially during stressed periods, while remaining reliable under simulated input degradation.
\end{abstract}


\section{Introduction}

Daily market-risk monitoring is often studied as a forecasting problem centered on conditional quantiles, Value-at-Risk (VaR) \cite{engle2004caviar}, and backtesting. Recent work extends this line with deep-learning approaches to tail-risk prediction \cite{fatouros2023deepvar}. Related work also explores foundation-model approaches for VaR forecasting \cite{goel2024foundationvar}. However, deployed monitoring services may fail for operational reasons even when the underlying predictive model appears statistically acceptable \cite{paleyes2022deploying}. In practice, degraded inputs and inconsistent records \cite{polyzotis2019data}, as well as missing macro data \cite{nist2026monitoring}, can make next-day downside-risk estimates unreliable in deployment. Serving-time distribution mismatches can further reduce predictive reliability \cite{ovadia2019trust}. This issue is especially relevant for Exchange-Traded Fund (ETF) surveillance, where daily monitoring may need to combine market information, cross-asset signals, and macro-financial indicators in a rolling surveillance process \cite{adrian2019monitoring}. The key problem is not only how to predict lower-tail risk, but also when the prediction should be trusted, qualified, or made more conservative. We therefore study ETF tail-risk surveillance primarily as a monitoring problem.

This paper makes three main contributions. First, we frame daily ETF tail-risk estimation as a service reliability problem rather than only a next-day VaR forecasting task. Second, we develop a monitoring architecture that combines input-quality diagnostics, uncertainty scoring, and a lower-tail adjustment rule, so that the reported VaR becomes more conservative when input quality deteriorates or predictive reliability weakens. Third, across multiple ETFs and under deliberate service-time input corruption, we show that this design improves stress-period breach control and service robustness while maintaining uninterrupted daily monitoring output.

\section{Related Work}

Related work most relevant to this paper lies in three strands. The first concerns lower-tail risk forecasting and market-risk measurement, especially VaR modeling, conditional quantiles, and risk-model backtesting. Foundational work includes regression quantiles \cite{koenker1978rq}, dynamic quantile-based VaR models such as CAViaR \cite{engle2004caviar}, and standard backtesting frameworks based on unconditional and conditional coverage diagnostics \cite{christoffersen1998evaluating}. More recent studies extend this literature through deep-learning methods for VaR forecasting \cite{wang2024deepquantile}, higher-moment information for VaR and expected-shortfall prediction \cite{le2024highermoments}, one-sided VaR recalibration under stressed states \cite{zhong2026proxyreliance}, and forward-looking or jump-sensitive downside-risk measures \cite{maheu2026variancejumps}. Despite these differences, the emphasis remains largely on forecast construction and post hoc risk evaluation.

A second strand emphasizes data quality and observability in analytical systems, showing that monitoring performance depends not only on the predictive model itself but also on data completeness, consistency, and timeliness \cite{wang1996beyondaccuracy}. A third strand studies predictive uncertainty and reliability under distribution shift and non-stationarity in machine learning systems \cite{wu2026oodtimeseries}. Related financial work further suggests that monitoring reliability can depend on richer market signals and early-warning indicators for systematic risk \cite{ciciretti2025earlywarning}. Taken together, these studies indicate that degraded inputs and regime shifts can weaken operational reliability even when average predictive performance appears acceptable.

Our paper differs in focus and design: rather than treating tail-risk forecasting, data quality, and predictive reliability as separate issues, we integrate them into a single monitoring framework for daily ETF surveillance under stress and degraded inputs.

\section{Framework and Method}

\begin{figure*}[!t]
\centering
\includegraphics[width=\textwidth, trim={0 6cm 0 1cm}, clip]{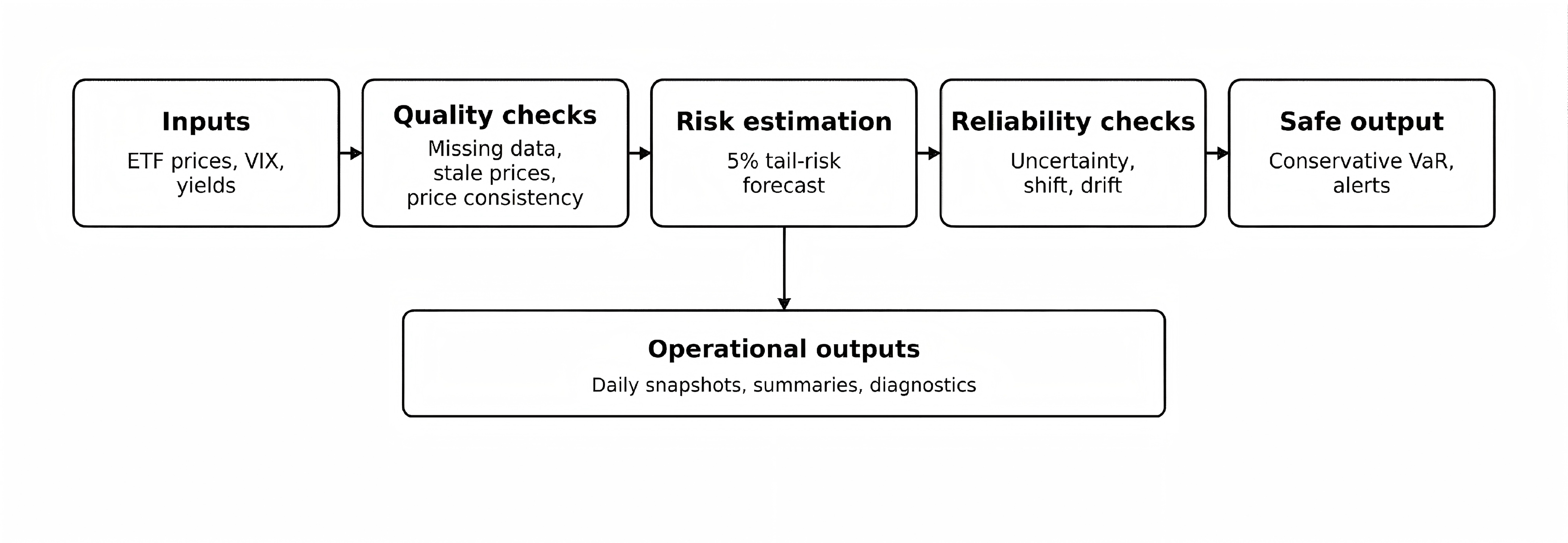}
\caption{Service pipeline of the proposed quality-aware and uncertainty-aware ETF risk monitoring framework.}
\label{fig:framework}
\end{figure*}

\subsection{Service Framework}
Fig.~\ref{fig:framework} summarizes the proposed monitoring pipeline. On each trading day $t$, the system receives ETF-panel inputs together with VIX and yield-curve information and proceeds through four stages: quality checks, risk estimation, reliability checks, and safe output generation. Each stage serves a separate role: input diagnosis, risk forecasting, reliability diagnosis, and conservative reporting.

\subsection{Data and Quality-Control Layer}

\subsubsection{Data Sources}

We use a daily panel of six ETFs spanning U.S.\ equities (SPY, QQQ, IWM), emerging markets (EEM), gold (GLD), and long-term Treasuries (TLT), augmented with VIX and zero-coupon yield information. ETF prices and trading volume are obtained from Bloomberg, while macro-financial series are obtained from FRED. For each asset-date observation, the main inputs include open, high, low, and close (OHLC) prices, volume, return, rolling volatility measures such as EWMA volatility, range-based proxies such as Parkinson volatility \cite{parkinson1980extreme} and Garman--Klass volatility \cite{garman1980estimation}, and drawdown-related variables such as cumulative peak price and drawdown.

\subsubsection{Quality Flags}
The quality-control layer assesses service-time input quality using only prediction-time observable variables. We distinguish between critical market fields and auxiliary macro inputs. The critical fields are \{open, high, low, close, volume, return\}. VIX and yield-curve variables are treated as auxiliary macro inputs: they are used when available, but missing macro values do not block daily output. We use a fixed service-layer parameterization with pre-specified windows, weights, and thresholds that are applied uniformly across assets. This design assesses whether a single stable configuration can improve monitoring reliability under degraded-input conditions.

Let $q_t^{\mathrm{miss}} \in [0,1]$ denote the fraction of missing values among the critical fields. Let $q_t^{\mathrm{ohlc}} \in \{0,1\}$ denote the OHLC-consistency flag. It equals 1 if the recorded OHLC values are internally inconsistent, and 0 otherwise. Let
$z^{r}_t$ denote the 60-day z-score of the current return and let
$z^{V}_t$ denote the 20-day z-score of the current volume. Define
\begin{equation}
\sigma(x;c,s)=\frac{1}{1+\exp\!\left(-(x-c)/s\right)}.
\end{equation}
Let $r_t$ denote the current-day return observed at prediction time $t$. We use two bounded anomaly scores: $q_t^{\mathrm{jump}} \in [0,1]$ for return anomalies and $q_t^{\mathrm{vol}} \in [0,1]$ for volume anomalies. They are given by
\begin{equation}
q_t^{\mathrm{jump}}
=
\max\!\left\{
\mathbf{1}\!\left(|r_t|>0.15\right),
\,
\sigma\!\left(|z_t^{r}|;3,1\right)
\right\},
\end{equation}
and
\begin{equation}
q_t^{\mathrm{vol}}
=
\sigma\!\left(|z_t^{V}|;3,1\right).
\end{equation}
Let $q_t^{\mathrm{stale}} \in \{0,1\}$ denote an indicator of at least two consecutive repeated close prices up to day $t$. These components are aggregated into the bounded quality score
\begin{equation}
\begin{aligned}
Q_t
={}&0.30\,q^{\mathrm{miss}}_t
+0.35\,q^{\mathrm{ohlc}}_t
+0.15\,q^{\mathrm{jump}}_t \\
&+0.10\,q^{\mathrm{vol}}_t
+0.10\,q^{\mathrm{stale}}_t.
\end{aligned}
\end{equation}
Larger values of $Q_t$ indicate poorer input quality. We map $Q_t$ into three input-quality states: green for $Q_t \le 0.25$, yellow for $0.25 < Q_t \le 0.60$, and red for $Q_t > 0.60$. The score $Q_t$ is used both as a predictive feature and in the conservative adjustment.

\subsection{Risk Model and Uncertainty Scoring}

\subsubsection{Prediction Target}
The risk-estimation stage produces a model-based next-day lower-tail estimate for the downstream layers. Let $r_{t+1}$ denote the next-day ETF return. The monitoring target is the conditional lower-tail threshold of $r_{t+1}$ at $\alpha=0.05$.

\subsubsection{Risk Model}
Let $X_t$ denote the feature vector observed at time $t$. We fit a single pooled panel model to all ETF-date observations, with symbol identity and date-level cross-asset information included in the feature set. The baseline lower-tail prediction is
\begin{equation}
\hat{q}^{\,\mathrm{raw}}_{t+1}=f(X_t),
\end{equation}
where $f(\cdot)$ is the ensemble-mean predictor from bootstrap quantile models fitted at $\alpha=0.05$ \cite{friedman2001gbm}. To reduce systematic miscalibration, we apply a rolling residual-based adjustment. The calibrated estimate is
\begin{equation}
\hat{q}_{t+1}=\hat{q}^{\,\mathrm{raw}}_{t+1}+c_t,
\end{equation}
where $c_t$ is estimated from the most recent 63-trading-day calibration window:
\begin{equation}
c_t = \operatorname{Quantile}_{\alpha} \left\{ r_{s+1}-\hat{q}^{\,\mathrm{raw}}_{s+1} :\, s \in \mathcal{C}_t \right\},
\end{equation}
where $\mathcal{C}_t$ denotes the 63 trading days immediately preceding the current prediction period.

\subsubsection{Uncertainty Scoring}\label{subsubsec:uncertainty_Scoring}
The uncertainty layer quantifies service-time reliability by combining ensemble dispersion, an out-of-distribution score, and a recent drift score. Using five bootstrap models, let $\hat{q}^{(b)}_{t+1}$, $b=1,\ldots,5$, denote the next-day lower-tail predictions. Let $s_t$ denote the local volatility reference scale for the same symbol on day $t$. In practice, we take $s_t$ to be the precomputed 20-day rolling-volatility feature when available and otherwise use the 20-day rolling standard deviation of returns. Here $\operatorname{sd}(\cdot)$ denotes the sample standard deviation. The model-dispersion component is
\begin{equation}
u^{\mathrm{model}}_t
=
\min\!\left\{
1,\,
\frac{1}{3}
\frac{\operatorname{sd}\!\left(
\hat{q}^{(1)}_{t+1},\ldots,\hat{q}^{(5)}_{t+1}
\right)}{s_t}
\right\}.
\end{equation}
Missing feature values are imputed using medians from the current training window. Let $d_t$ denote the current PCA-based Mahalanobis distance \cite{lee2018mahalanobis}, computed in the corresponding PCA space, and let $d_t^{\mathrm{ref}}$ denote the 95th percentile of the in-training distances. Define
\begin{equation}
u^{\mathrm{ood}}_t
=
\mathrm{clip}\!\left(
\frac{d_t / d_t^{\mathrm{ref}} - 1}{1.5},
\,0,\,1
\right),
\end{equation}
where $\mathrm{clip}(x,0,1)=\min\{\max(x,0),1\}$. Let $\hat{p}^{\,\mathrm{post}}_t$ denote the empirical breach rate for the same symbol over its most recent 60 post-adjustment forecasts. Then
\begin{equation}
u^{\mathrm{drift}}_t
=
\min\!\left\{
1,\,
\frac{\max(\hat{p}^{\,\mathrm{post}}_t-\alpha,0)}{2\alpha}
\right\},
\end{equation}
with $u^{\mathrm{drift}}_t=0$ if fewer than 30 observations are available. The final uncertainty score is
\begin{equation}
U_t
=
0.40\,u^{\mathrm{model}}_t
+
0.35\,u^{\mathrm{ood}}_t
+
0.25\,u^{\mathrm{drift}}_t.
\end{equation}
The score $U_t$ scales the conservative adjustment and also defines a two-state uncertainty label for diagnostic reporting. Let
\begin{equation}
\tau_t^{U}
=
\operatorname{Quantile}_{0.90}
\left\{
U_s : s \in \mathcal{H}^{U}_t
\right\},
\end{equation}
where $\mathcal{H}^{U}_t$ contains the most recent 252 evaluable prediction dates for the same symbol before day $t$, with shorter histories used only at the beginning of the sample. The uncertainty state is low if $U_t \le \tau_t^{U}$ and elevated otherwise.

\subsection{Output Adjustment and Alerts}
The safe-output stage combines the calibrated estimate $\hat{q}_{t+1}$ with a 63-day historical VaR benchmark $\hat{q}^{\,\mathrm{hist},63}_{t+1}$, defined as the empirical 5\% quantile of the previous 63 trading days of returns for the same symbol. The final safe VaR is
\begin{equation}
\hat{q}^{\,\mathrm{safe}}_{t+1}
=
\min\!\left\{
\hat{q}^{\,\mathrm{hist},63}_{t+1},
\hat{q}_{t+1}-A_t
\right\},
\end{equation}
where
\begin{equation}
A_t
=
s_t\bigl(0.75\,U_t + 0.50\,Q_t\bigr),
\end{equation}
and $s_t$ is the same local volatility reference scale defined earlier. We also define the fallback-intensity ratio
\begin{equation}
R_t = \frac{\bigl(\hat{q}_{t+1}-\hat{q}^{\,\mathrm{safe}}_{t+1}\bigr)_+}{s_t},
\end{equation}
where $(x)_+=\max(x,0)$. For alerts, we use a three-level uncertainty label: low for $U_t \le 0.33$, medium for $0.33 < U_t \le 0.66$, and high for $U_t > 0.66$. A red alert is issued if the quality state is red, the uncertainty label is high, $u_t^{\mathrm{drift}} \ge 1.0$, or $R_t \ge 0.75$; an orange alert is issued if the quality state is yellow, the uncertainty label is medium, $u_t^{\mathrm{drift}} \ge 0.5$, or $R_t \ge 0.35$; all remaining cases are labeled green. All windows, weights, and thresholds are fixed ex ante and kept unchanged across experiments.

\section{Experimental Design}

\subsection{Backtest Protocol}

We evaluate the proposed monitoring service under a rolling walk-forward design that mimics daily deployment. The ensemble is refit every 63 trading days using the most recent 756 trading days of training data, and a residual-based adjustment is calibrated from the immediately preceding 63 trading days. Daily predictions are then issued until the next refit. The drift component is updated separately from the same symbol's most recent 60 evaluable post-adjustment predictions. The prediction period runs from 2023-01-03 to 2025-12-29.

Future returns are used only for evaluation, so the backtest remains aligned with a realistic daily monitoring workflow.

\subsection{Model Specification}
The predictive module is a 5\% quantile gradient-boosting ensemble, obtained by averaging predictions from multiple gradient-boosting models trained on date-level bootstrap resamples of the training window. A residual-based quantile adjustment estimated from the 63-trading-day calibration window is then added to the ensemble-mean prediction.

The uncertainty layer combines three signals: ensemble dispersion \cite{lakshminarayanan2017deepensembles}, a PCA--Mahalanobis out-of-distribution score that measures distance from the training distribution, and a recent drift score based on the same asset's recent breach frequency. These are aggregated into a single uncertainty score used both for conservative adjustment and for the low/elevated diagnostic split.

\subsection{Comparison Baselines and Proposed Service Output}

We compare the proposed service output against four benchmark risk measures: the unconstrained model prediction before fallback, a 252-day historical VaR benchmark, an EWMA-normal VaR benchmark, and a GJR-GARCH(1,1) VaR benchmark. The 252-day historical VaR is computed as the empirical 5\% quantile of the previous 252 trading days of returns. The EWMA-normal benchmark is constructed from the precomputed EWMA volatility feature in the input panel, following the RiskMetrics-style EWMA approach \cite{jpmorgan1996riskmetrics}. If $\sigma^{\mathrm{ewma}}_t$ denotes that feature, then
\begin{equation}
\hat{q}^{\,\mathrm{ewma}}_{t+1}
=
-1.64485\,\sigma^{\mathrm{ewma}}_t,
\end{equation}
which is the Gaussian 5\% lower-tail threshold. The GJR-GARCH benchmark follows Glosten, Jagannathan, and Runkle \cite{glosten1993gjr}, uses Student-\emph{t} innovations, is re-estimated at each retrain point, and is rolled forward between retraining dates for one-step-ahead VaR evaluation. It therefore follows the same 63-trading-day retraining schedule as the proposed predictive module.

The final safe VaR combines the model-based estimate with the shorter-window benchmark $\hat{q}^{\,\mathrm{hist},63}_{t+1}$ and an uncertainty- and quality-driven adjustment. Unlike the 252-day historical VaR, EWMA-normal VaR, and GJR-GARCH VaR, $\hat{q}^{\,\mathrm{hist},63}_{t+1}$ is used only as an internal fallback anchor.

\subsection{Quality-Validation Experiment}
To complement the main clean-sample backtest, we conduct an additional validation experiment based on simulated input degradation \cite{zhou2024dqsurvey}. Lightweight faults are injected only at prediction time into service-time observable inputs using a nominal row-wise corruption probability of 15\%. Three fault modes are considered: missing critical fields, stale-price patterns, and OHLC inconsistencies. Missing-mode corruption removes two to four raw service-time fields, stale-mode corruption flattens same-day prices, and OHLC-mode corruption introduces invalid price relationships.

We compare three variants under the same corrupted inputs: the full service, a version without the quality feature, and a version without quality-based service actions. In the third variant, the predictive model still uses the quality feature, but quality-based fallback adjustment and alert escalation are disabled. This experiment tests whether the quality layer adds operational value under degraded inputs through both prediction and downstream actions.

\subsection{Evaluation Metrics}
We evaluate the framework from predictive and operational perspectives. Predictive performance is assessed by empirical breach rate, defined as the fraction of cases in which the realized return falls below the corresponding VaR threshold, the Kupiec likelihood-ratio test for unconditional coverage \cite{kupiec1995var}, and pinball loss \cite{fissler2016higher}. We report these metrics for the unconstrained model output, the proposed service output (Safe VaR), and the baseline benchmarks, emphasizing unconditional reliability and operational robustness rather than a full set of dynamic backtests.

We also examine stress-period reliability, performance across the low and elevated uncertainty states defined in Subsubsection~\ref{subsubsec:uncertainty_Scoring}, cross-asset stability through ETF-level breach-rate comparisons, and operational behavior through alert frequencies and output availability. Stress periods are defined as dates for which the VIX is at or above the 80th percentile of the pooled VIX distribution, independently of model outcomes.

\section{Results}

\subsection{Main Monitoring Performance}
Table~\ref{tab:main_results} reports the main predictive and service-level results on the prediction period. The unconstrained model VaR attains an overall breach rate of 5.80\%, indicating mild under-coverage relative to the 5\% target. After fallback, Safe VaR reduces the breach rate from 5.80\% to 4.35\%, with the reported tail-risk estimate adjusted downward relative to the unconstrained model. The 252-day historical VaR benchmark is also conservative at 4.42\%, whereas the EWMA-normal benchmark remains under-covered at 5.75\%. The GJR-GARCH(1,1)-\emph{t} benchmark achieves a breach rate of 5.31\% with the lowest Kupiec statistic (0.89) and the lowest pinball loss among all methods, confirming that it is a competitive volatility-based risk model. However, as we show below, its coverage degrades under stress and varies substantially across assets.

\begin{table}[!htbp]
\caption{Main predictive and service-level results on the prediction period.}
\label{tab:main_results}
\centering
\setlength{\tabcolsep}{3.5pt}
\footnotesize
\begin{tabular}{@{}lccc@{}}
\toprule
Method & Breach rate (\%) & Kupiec LR & Pinball loss \\
\midrule
Model VaR                  & 5.80 & 5.76 & 0.001255 \\
Safe VaR                   & 4.35 & 4.12 & 0.001251 \\
Historical VaR (252-day)   & 4.42 & 3.30 & 0.001285 \\
EWMA-normal VaR            & 5.75 & 5.15 & 0.001242 \\
GJR-GARCH(1,1)-\emph{t} VaR & 5.31 & 0.89 & 0.001235 \\
\bottomrule
\end{tabular}
\end{table}

The Kupiec results show more specifically how conservative fallback changes unconditional coverage. Relative to the unconstrained model, the safe output reduces the breach rate and lowers the Kupiec statistic from 5.76 to 4.12, but it remains in the 5\% rejection region under the Kupiec test. The safe output should therefore be interpreted as a partial improvement in unconditional coverage rather than a full pass of the Kupiec criterion.

\begin{figure}[htp]
\centering
\includegraphics[width=\columnwidth]
{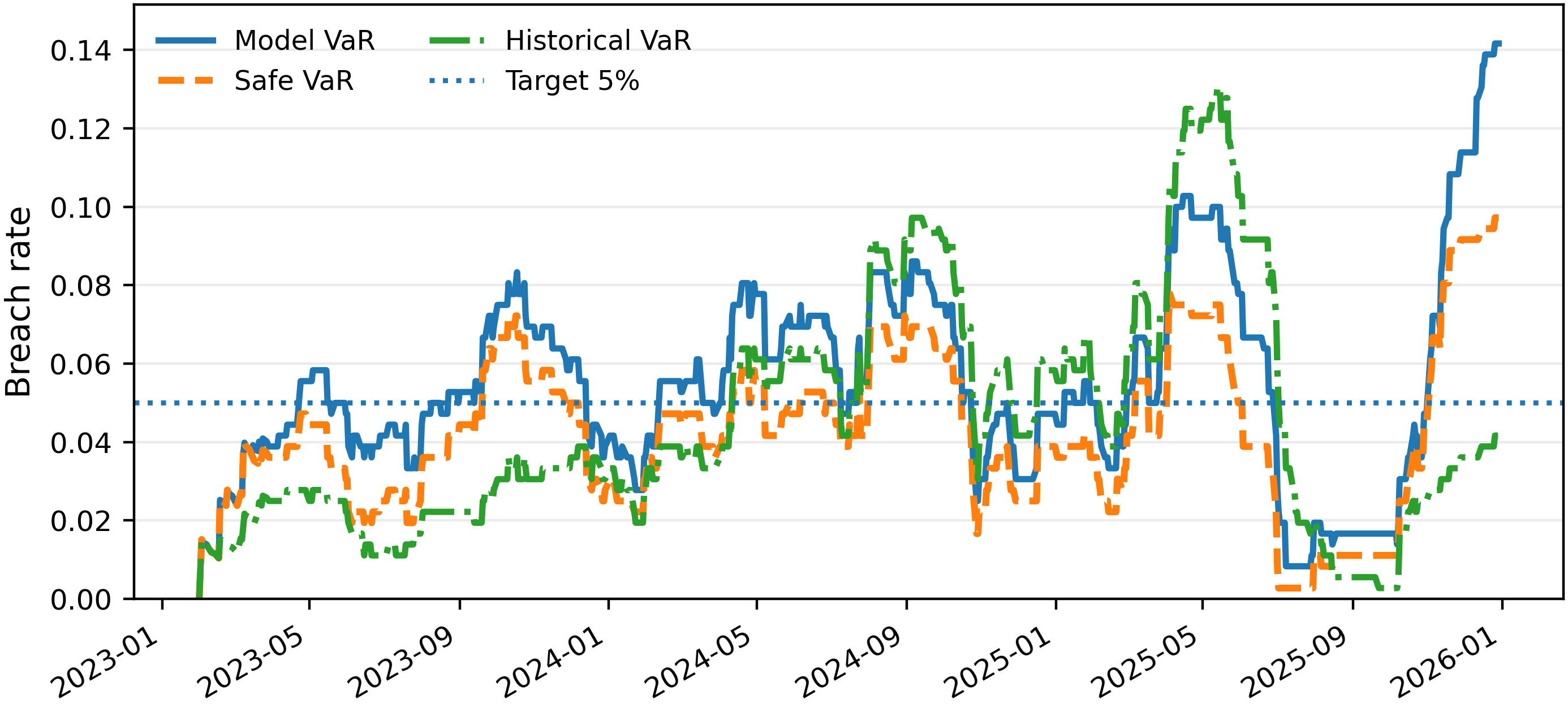}
\caption{Rolling 60-day breach rates comparison.}
\label{fig:rolling_breach}
\end{figure}

Fig.~\ref{fig:rolling_breach} shows the 60-day rolling breach rates. The safe output is more stable over time than the unconstrained model, while the 252-day historical VaR benchmark also becomes unstable in more volatile parts of the sample.

\subsection{Reliability During High-VIX Periods}

\begin{figure}[htp]
\centering
\includegraphics[width=\linewidth]{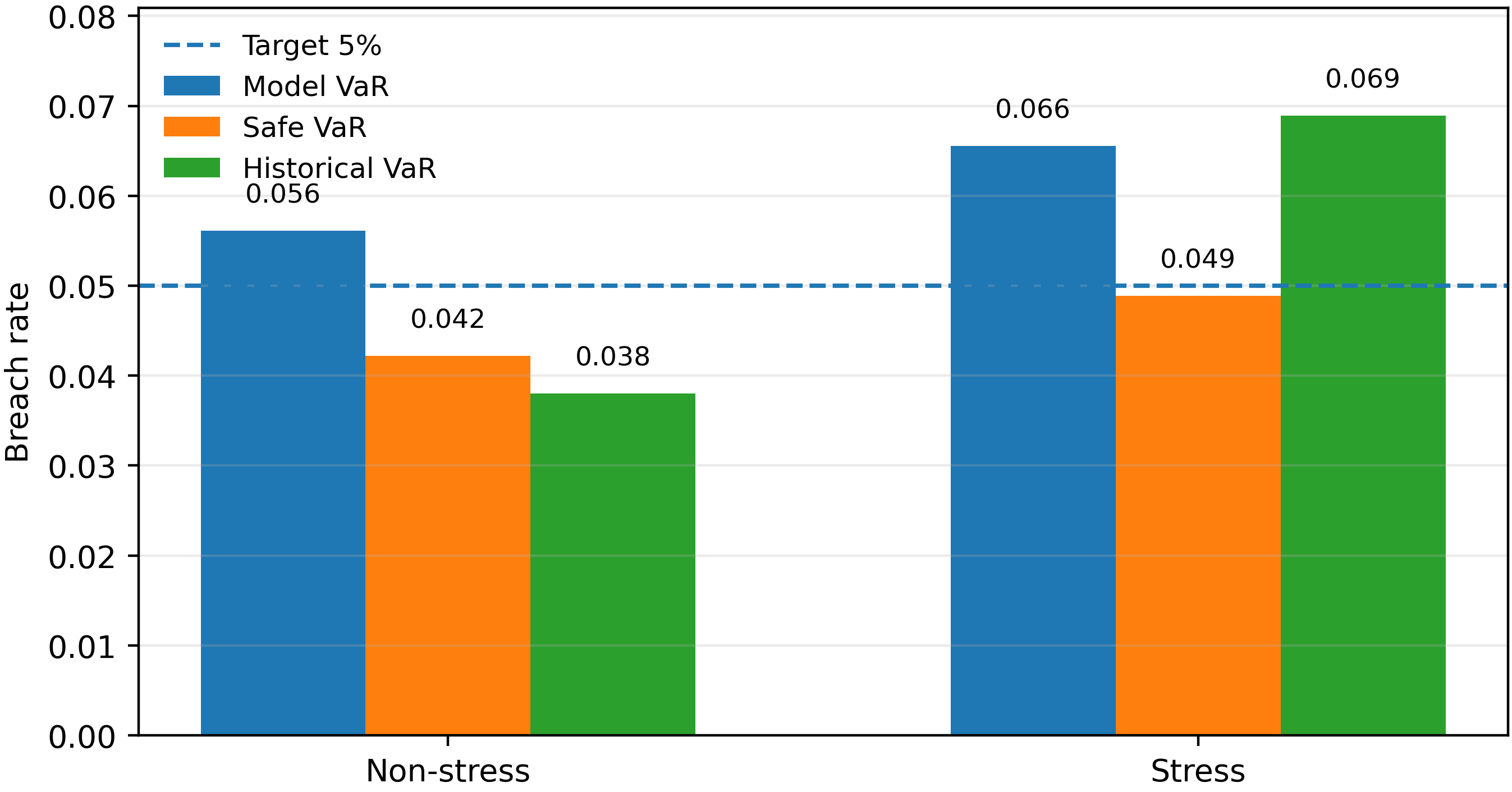}
\caption{Breach rates in non-stress and stress regimes.}
\label{fig:stress_compare}
\end{figure}

Fig.~\ref{fig:stress_compare} compares breach rates in non-stress and stress regimes. In non-stress periods, the safe output is somewhat conservative but remains close to the target. Under stress, its advantage becomes clearer: the safe layer stays nearer the target level, whereas the unconstrained model and the simpler benchmark rules exhibit higher breach rates. This pattern indicates that the fallback layer is particularly valuable when market conditions deteriorate; even the stronger GJR-GARCH baseline has a higher stress-period breach rate (5.78\%) than Safe VaR (4.89\%).

\subsection{Component Contributions to Safe VaR}

To clarify whether the gains come mainly from a simple conservative rule or from the full design, we compare five variants: the raw model, a simple historical fallback, quality-only fallback, uncertainty-only fallback, and the full service. The simple fallback uses only the 63-day historical anchor. The quality-only and uncertainty-only variants retain that same anchor but use only one adjustment component, whereas the full service combines both.

\begin{table}[htp]
\caption{Component comparison of safe-fallback breach rates (\%).}
\label{tab:fallback_components}
\centering
\scriptsize
\setlength{\tabcolsep}{4pt}
\begin{tabular}{lcccc}
\toprule
& \multicolumn{2}{c}{Clean inputs} & \multicolumn{2}{c}{Corrupted inputs} \\
\cmidrule(lr){2-3}\cmidrule(lr){4-5}
Variant & Overall & Stress & Overall & Stress \\
\midrule
Raw model                 & 5.80 & 6.56 & 5.78 & 6.56 \\
Simple fallback           & 4.64 & 5.56 & 4.67 & 5.56 \\
Quality-only              & 4.55 & 5.33 & 4.49 & 5.33 \\
Uncertainty-only          & 4.47 & 5.00 & 4.44 & 4.89 \\
Full service              & 4.35 & 4.89 & 4.24 & 4.67 \\
\bottomrule
\end{tabular}
\end{table}

Table~\ref{tab:fallback_components} reports breach rates for the five variants on the original clean-input sample and on the corrupted-input sample, each evaluated both over all prediction dates and over the stress subset only. The ranking is consistent across all four settings. All safeguarded variants improve on the raw model, and the full service performs best throughout. The uncertainty-only fallback is consistently closer to the full service than the quality-only fallback, indicating that uncertainty-aware adjustment is the main source of the gain, while the quality layer provides additional protection under degraded inputs.

\subsection{Performance Across Uncertainty States}
Table~\ref{tab:uncertainty_results} reports Safe VaR performance across two diagnostic uncertainty states. For each symbol, the state is labeled as elevated when the current uncertainty score exceeds the rolling 90th percentile of prior uncertainty scores, and as low otherwise.

The split is directionally consistent: most prediction days fall into the low-uncertainty state, whereas the elevated state has higher breach rates and larger pinball loss. Higher uncertainty is associated with weaker monitoring performance.

\begin{table}[htp]
\caption{Safe-output performance across low and elevated uncertainty states.}
\label{tab:uncertainty_results}
\centering
\small
\begin{tabular}{lccc}
\toprule
Uncertainty state & Count & Breach rate (\%) & Pinball loss \\
\midrule
Low      & 3839 & 4.27 & 0.001198 \\
Elevated &  662 & 4.83 & 0.001557 \\
\bottomrule
\end{tabular}
\end{table}

\subsection{Cross-Asset Stability and Service Availability}

Table~\ref{tab:asset_results} reports safe-output and GJR-GARCH breach rates by ETF. The safe service remains reasonably stable across the six monitored symbols, with breach rates below 5\% for EEM, GLD, IWM, SPY, and TLT; the most difficult case is QQQ at 5.73\%. By contrast, the GJR-GARCH baseline exceeds 5\% on three of six ETFs (EEM at 7.60\%, QQQ at 6.40\%, and SPY at 5.19\%), showing that its overall coverage appears acceptable, but its breach rates vary substantially across assets. Taken together, these results suggest that the proposed service is more stable across assets than the external econometric benchmark.

At the service level, the framework generates 4501 evaluable predictions, for 100\% output availability. Macro curve information is available on 88.9\% of prediction days. The alert distribution is 3612 green, 726 orange, and 163 red, indicating continuous output generation with interpretable alert behavior.

\begin{table}[htp]
\caption{Cross-asset breach rates.}
\label{tab:asset_results}
\vspace{-2mm}
\centering
\scriptsize
\setlength{\tabcolsep}{10pt}
\renewcommand{\arraystretch}{0.95}
\begin{tabular}{@{}lcc|lcc@{}}
\toprule
ETF & Safe (\%) & GJR (\%) & ETF & Safe (\%) & GJR (\%) \\
\midrule
EEM & 4.80 & 7.60 & QQQ & 5.73 & 6.40 \\
GLD & 3.07 & 4.40 & SPY & 4.39 & 5.19 \\
IWM & 4.80 & 3.87 & TLT & 3.33 & 4.40 \\
\bottomrule
\end{tabular}
\vspace{-3mm}
\end{table}

\subsection{Quality-Layer Validation Under Input Degradation}
We finally examine a focused quality-layer ablation under simulated service-time input corruption. Table~\ref{tab:quality_validation} compares the full service with two ablations: a model-side ablation that removes the quality score from the predictive feature set, and a service-side ablation that retains the quality feature in prediction but disables quality-based fallback adjustment and alert escalation.

The overall breach rate for the full service is 4.24\%, versus 4.27\% without the quality feature and 4.44\% without the quality service layer. In stress periods, the full service again attains the lowest breach rate at 4.67\%, whereas both ablations are higher at 4.89\%. Pinball-loss differences are small, suggesting that under corrupted inputs the quality layer contributes mainly to robustness rather than sharpness.

Disabling quality-based service actions mainly shifts alerts from orange to green, with little effect on the most severe alerts.

\begin{table}[htp]
\caption{Focused quality-layer ablation under simulated service-time input degradation. Alerts (G/O/R) denote the counts of green, orange, and red alerts.}
\label{tab:quality_validation}
\centering
\small
\resizebox{\columnwidth}{!}{%
\begin{tabular}{lcccc}
\toprule
Experiment & Overall (\%) & Stress (\%) & Pinball & Alerts (G/O/R) \\
\midrule
Corrupt full service        & 4.24 & 4.67 & 0.001254 & 3472/875/154 \\
No quality feature          & 4.27 & 4.89 & 0.001258 & 3443/897/161 \\
No quality service layer    & 4.44 & 4.89 & 0.001252 & 3679/668/154 \\
\bottomrule
\end{tabular}%
}
\end{table}

\section{Conclusion and Future Work}

Empirically, the proposed framework improves stress-period breach control, remains stable across assets, and maintains full output availability throughout the prediction period. These results suggest that practical ETF monitoring must consider not only forecast accuracy, but also reliable operation under degraded data, market shifts, and higher uncertainty.

Several limitations remain. The study uses a relatively small ETF sample, and the isolated contribution of the quality layer is evaluated mainly through simulated service-time degradation experiments. In addition, the service-layer weights and thresholds are fixed design choices and are not supported by a broader sensitivity analysis. The framework also operates at daily rather than intraday frequency, and some alert and fallback components still leave room for further optimization. Future work can extend the analysis to a broader set of assets, incorporate richer uncertainty and regime-shift diagnostics, and connect monitoring outputs to downstream decision rules such as hedging, capital allocation, and alert escalation.

\smallskip

\noindent\textbf{Code availability:} Code and full hyperparameters are available at \href{https://github.com/TenghanZhong/etf-tail-risk-monitoring}{github.com/TenghanZhong/etf-tail-risk-monitoring}.

\bibliographystyle{IEEEtran}
\bibliography{references_verified}

\end{document}